\documentstyle[aps,prb,epsfig,multicol]{revtex}

\begin{document}

\draft

\title{%
Friedel Oscillations in the Open Hubbard Chain
}
\author{G.~Bed\"urftig}
\address{Institut f\"ur Theoretische Physik,
Universit\"at Hannover, D-30167 Hannover, Germany}
\author{B.~Brendel}
\address{Institut f\"ur Theoretische Physik, 
Universit\"at W\"urzburg, Am Hubland, D-97074 W\"urzburg, Germany}
\author{H.~Frahm}
\address{Institut f\"ur Theoretische Physik,
Universit\"at Hannover, D-30167 Hannover, Germany}
\author{R.M.~Noack}
\address{Institut de Physique Th\'eorique, Universit\'e de Fribourg,
CH-1700 Fribourg, Switzerland}
\date{\today}
\maketitle
\begin{abstract}
  Using the Density Matrix Renormalization Group (DMRG), we calculate
  critical exponents for the one-dimensional Hubbard model with open
  boundary conditions with and without additional boundary potentials at
  both ends.  A direct comparison with open boundary condition Bethe Ansatz
  calculations provides a good check for the DMRG calculations on large
  system sizes.  On the other hand, the DMRG calculations provide an
  independent check of the predictions of Conformal Field Theory, which are
  needed to obtain the critical exponents from the Bethe Ansatz.  From
  Bethe Ansatz we predict the behaviour of the $1/L$-corrected mean value
  of the Friedel oscillations (for the density and the magnetization) and
  the characteristic wave vectors, and show numerically that these
  conjectures are fulfilled with and without boundary potentials.  The
  quality of the numerical results allows us to determine, for the first
  time, the behaviour of the coefficients of the Friedel oscillations as a
  function of the the Hubbard interaction.
\end{abstract}

\ifpreprintsty\onecolumn\fi

\makeatletter
\global\@specialpagefalse
\def\@oddhead{\hfill ITP-UH-10/98 and WUE-ITP-98-015}
\let\@evenhead\@oddhead
\makeatother

\begin{multicols}{2}


\section{Introduction}
\label{sec:Intro+Model}

Recent Bethe Ansatz studies of the one-dimensional Hubbard model with open
boundaries subject to boundary chemical potentials or magnetic fields
\cite{hschulz:85,assu:96,deyu:pp,shwa:97} have opened new possibilities to
apply the predictions of Boundary Conformal Field Theory
\cite{card:89,aflu:94,affl:94} for the asymptotics of correlation functions
to quantum impurity problems.  As in the case of periodic systems,
\cite{frko:90,frko:91,kaya:91} the corresponding matrix elements cannot be
computed directly but must be extracted from the scaling behaviour of the
low--lying excited states.  For generic filling and magnetization these
finite size spectra allow the identification of the contributions from two
massless bosonic sectors associated with the spin and charge excitations in
a Tomonaga-Luttinger liquid.

A crucial step in these studies of systems with open boundary conditions is
the correct interpretation of the finite-size spectra obtained from the
Bethe Ansatz.  These spectra determine both the bulk correlation functions,
which are independent of the boundary fields, and the nonuniversal
dependence of boundary phenomena such as the orthogonality exponent or
X-ray edge singularities on the strength of the scatterer.
Hence both in the requirement of conformal invariance and in the analysis
of the finite-size spectra, the computation of correlation functions relies
on assumptions which need to be verified by a more direct method, most
notably by numerical calculations.  Of course, comparison with exact
results is also of interest for the numerical calculation: predictions of
critical exponents for microscopic models allow the algorithms to be
improved, which can then, in turn, be expected to produce better and more
reliable results for more general systems.

These considerations motivate our study of the Friedel oscillations for the
single particle density and magnetization in the open Hubbard chain with
boundary chemical potentials, which is described by the Hamiltonian
\begin{eqnarray}
{\cal H} = &-& \sum_{\sigma,j=1}^{L-1} \left( c_{j,\sigma}^\dagger
    c_{j+1,\sigma} + {\rm h.c.} \right) + U \sum_{j=1}^L n_{j,\uparrow}
  n_{j,\downarrow} \nonumber \\
   &-& p \left( n_{1,\uparrow} +
     n_{1,\downarrow}+n_{L,\uparrow} +
     n_{L,\downarrow} \right)  ,
\label{ham:hubb}
\end{eqnarray}
where the lattice has $L$ sites, $c_{j,\sigma}^\dagger$ ($c_{j,\sigma}$)
creates (destroys) an electron on site $j$, $n_{j,\sigma} =
c_{j,\sigma}^\dagger c_{j,\sigma}$, and we have made the hopping integral
dimensionless so that the Coulomb repulsion $U$ and the on--site potential
$p$ are measured in dimensionless units.

Numerical calculations of critical exponents in low dimensional systems
such as magnetic chains or electronic systems have a long history \cite{}.
Due to the need to consider systems of sufficient size, many earlier
studies have used Quantum Monte Carlo methods to treat systems with
periodic boundary conditions (PBC). \cite{sopa:90} More recently, the
Density Matrix Renormalization Group (DMRG) \cite{whit.92,whit.93} has
become a new and powerful method especially suited to the study of one
dimensional systems with open boundary conditions (OBC).
\cite{hifu:98,meschmi:98} However systems with PBC have also been studied
with this method. \cite{qili:95,schmeck:96,shueni:96}

In the usual approach to the calculation of correlation functions with DMRG
one considers large systems and averages over an ensemble of correlation
functions located sufficiently far from the boundaries.  In the
thermodynamic limit, this procedure removes the Friedel oscillations due to
the boundary and gives the bulk behaviour of the quantity in question.

Here we want to make use of the existence of the exact solution of Eq.\ 
(\ref{ham:hubb}) in two ways in order to prepare the way for further
extensions of the method: First, we use the quantities obtained from the
Bethe Ansatz to provide checks of the numerical method at large system
sizes.  Second, we use the information contained in the oscillating
behaviour to obtain more reliable results for the critical exponents.

This paper is organized as follows.  In Sec.\ \ref{sec:Meth}, we give a
short description of the Bethe Ansatz and DMRG methods, citing the relevant
results from the Bethe Ansatz and Conformal Field Theory (CFT) concerning
Friedel oscillations.
In Sec.\ \ref{sec:fr_osc}, we study the Friedel oscillations of the density
$N(x)$ and the magnetization $M(x)$.  Combining the CFT results with those
for noninteracting fermions, we obtain conjectures for the explicit form of
the Friedel oscillations.  After introducing the fit method used to obtain
the exponents and coefficients from the DMRG results, we check the
conjectures for two fixed densities and varying on--site interaction $U$.
In addition, we study the dependence of the exponents and coefficients on
the boundary potential $p$ at one density.


\section{Methods}
\label{sec:Meth}


\subsection{Bethe Ansatz}
\label{sec:BA}

The one-dimensional Hubbard model with OBC, Eq.\ (\ref{ham:hubb}), can be
solved using the coordinate Bethe ansatz.  The symmetrized Bethe Ansatz
equations (BAE) determining the spectrum of ${\cal H}$ in the
$N_{e}=N_\uparrow+N_\downarrow$-particle sector read
\cite{hschulz:85,assu:96,deyu:pp,shwa:97}
\begin{eqnarray}
  e^{ik_j 2L} B_c(k_j) = 
  \prod_{\beta=-N_\downarrow}^{N_\downarrow}
  \frac{\sin(k_j)-\lambda_\beta+iu}{\sin(k_j)-\lambda_\beta-iu} \nonumber
  \\ \nonumber \\
  B_s(\lambda_\alpha) \prod_{j=-N_e}^{N_e}  
  \frac{\lambda_\alpha-\sin(k_j)+iu}{\lambda_\alpha-\sin(k_j)-iu} = 
  \prod_{\beta=-N_\downarrow \atop \beta \not= \alpha}^{N_\downarrow}  
  \frac{\lambda_\alpha-\lambda_\beta+2iu}{\lambda_\alpha-\lambda_\beta-2iu}
  \nonumber  \\ 
  j=-N_e,\ldots,N_e  \quad;\quad
  \alpha=-{N_\downarrow},\ldots,{N_\downarrow}  
  \label{eq:bae}
\end{eqnarray}
where we have defined $u=U/4$ and identified the solutions $k_{-j}=-k_j$
and $\lambda_{-\alpha}=-\lambda_{\alpha}$ in order to simplify the BAE.
The boundary terms read
\begin{eqnarray}
  B_c(k)&=&\left(\frac{e^{ik}-p}{1-p e^{ik}}\right)^2 
  \frac{\sin(k)+iu}{\sin(k)-iu} \,,\nonumber \\
  B_s(\lambda)&=& 
  \frac{\lambda+2iu}{\lambda-2iu} \,.
  \label{bound}
\end{eqnarray}

Since the BAE are already symmetrized and the solutions $k=0$ and
$\lambda=0$ have to be excluded, the energy of the corresponding eigenstate
of Eq.\ (\ref{ham:hubb}) is given by
\begin{equation}
  E=1- \sum_{j=-N_e}^{N_e} \cos{}(k_j) .
  \label{energy}
\end{equation}
In Refs. \onlinecite{assu:96,deyu:pp,shwa:97} the ground state and the
low--lying excitations were studied for small boundary fields.  In Ref.
\onlinecite{befr:97} the existence of boundary states in the ground state
for $p>1$ was established.  Bound states occur as additional complex
solutions for the charge and spin rapidities.

Here we will use the explicit form of the BAE, Eq.\ (\ref{eq:bae}), to
check the energy convergence of the DMRG results for finite $L$.
Furthermore, the expectation values of the density at the boundaries can be
calculated from the derivative of the energy with respect to $p$ (cf.\ 
Sec.\ \ref{sec:DMRG}) allowing another check of the numerics.  Finally, the
value of the magnetization at the boundaries for vanishing $p$ can be
calculated with a slightly modified Bethe--Ansatz (i.e. with a magnetic
field at the boundary, see Refs. \onlinecite{deyu:pp,shwa:97}).

Using standard procedures, the BAE for the ground state and low--lying
excitations can be rewritten as linear integral equations for the densities
$\rho_c(k)$ and $\rho_s(\lambda)$ of real quasi-momenta $k_j$ and spin
rapidities $\lambda_\alpha$, respectively:
\begin{equation}
   \left( \begin{array}{c} \rho_c(k) \\[5pt] 
                           \rho_s(\lambda) \end{array} \right) =
    \left( \begin{array}{c} {1 \over \pi}+{1 \over L}\hat{\rho}_c^0(k)\\[5pt] 
                       {1 \over L}\hat{\rho}_s^0(\lambda) \end{array} \right)
    + K * \left( \begin{array}{c} \rho_c(k') \\[5pt] 
                           \rho_s(\lambda') \end{array} \right)\ 
\label{eq:dnorm}
\end{equation}
with the kernel $K$ given by
\begin{equation}
 K=
   \left( \begin{array}{cc} 0 & \cos{}(k)\ a_{2u}(\sin(k)-\lambda') \\[5pt] 
                            a_{2u}(\lambda-\sin(k')) & 
                -a_{4u}(\lambda-\lambda') \end{array} \right).
 \label{eq:kernel}
\end{equation}
Here we have introduced $a_y(x)={1 \over 2\pi}\frac{y}{y^2/4+x^2}$, and
$f*g$ denotes the convolution $\int_{-A}^{A}dy f(x-y)g(y)$ with boundaries
$A=k_0$ in the charge and $A=\lambda_0$ in the spin sector.  The values of
$k_0$ and $\lambda_0$ are fixed by the conditions
\begin{equation}
   \int_{-k_0}^{k_0}dk \rho_c =
        \frac{2\left[N_e-C_c\right]+1}{L}\quad
\label{eq:fix1}
\end{equation}
and
\begin{equation}
   \int_{-\lambda_0}^{\lambda_0}d\lambda\rho_s =
        \frac{2\left[N_\downarrow-C_s\right]+1}{L} ,
\label{eq:fix2}
\end{equation}
where $C_c$ $(C_s)$ denotes the number of complex $k$
$(\lambda)$--solutions present in the ground state.\cite{befr:97} In
addition to the boundary terms in Eq.\ (\ref{bound}), the driving terms
$\hat{\rho}_c^0$ and $\hat{\rho}_s^0$ depend on whether or not the complex
solutions are occupied or not.  The explicit form can be found in Refs.
\onlinecite{assu:96,deyu:pp,shwa:97,befr:97}.  The presence of these $1/L$
corrections leads to the shifts
\begin{equation}
  \theta^c_p = {1 \over 2}\left(L \int_{-k_0}^{k_0}dk \hat{\rho}_c-1+2
    C_c \right)
\label{eq:tpc1}
\end{equation}
\begin{equation}
\theta^s_p = {1 \over 2}\left(L \int_{-\lambda_0}^{\lambda_0}d\lambda 
  \hat{\rho}_s -1+2 C_s \right) ,
\label{eq:tpc2}
\end{equation}
where $\hat{\rho}_c$ and $\hat{\rho}_s$ denote the solution of Eq.\ 
(\ref{eq:dnorm}) without the $1/\pi$ driving term, i.e.\ the bulk system
solution.

Here we will be mainly interested in the exponents of the Friedel
oscillations, given in Table\ \ref{tab:gamma}.  The quantity that
determines the critical exponents is the dressed charge matrix ${\bf Z}$:
\cite{frko:90,woyn:89}
\begin{equation}
{\bf Z}= \left( \begin{array}{cc} Z_{cc} & Z_{cs} \\ 
  Z_{sc} & Z_{ss} \end{array} \right)=
  \left( \begin{array}{cc} \xi_{cc}(k_0) & \xi_{sc} (k_0) \\ 
  \xi_{cs} (\lambda_0) & \xi_{ss}(\lambda_0) \end{array} \right)^\top
\label{eq:z}
\end{equation}
which is defined in terms of the integral equation
\begin{equation}
\left( \begin{array}{cc} \xi_{cc}(k) & \xi_{sc}(k) \\ 
  \xi_{cs}(\lambda) & \xi_{ss}(\lambda) \end{array} \right)=
\left( \begin{array}{cc} 1 & 0 \\ 0 & 1 \end{array} \right)+
        K^\top *
 \left( \begin{array}{cc} \xi_{cc}(k') & \xi_{sc}(k') \\ 
  \xi_{cs}(\lambda') & \xi_{ss}(\lambda') \end{array} \right) \,. \ 
\end{equation}
In Ref. \onlinecite{card:84} it was shown that the $n$-point correlation
functions of the open boundary system are related to the $2n$-point
functions of the periodic boundary system.  Thus the expectation value of
the local density in the open system, $\langle N(x) \rangle_o$ can be
extracted from the two--point correlation function $\langle N(z_1) N(z_2)
\rangle_p$ of the periodic system (see also Ref.\ \onlinecite{wavp:96}
where a spinless fermion model was considered).  We can therefore use the
results obtained in Refs. \onlinecite{frko:90} and \onlinecite{frko:91} for
the density--density correlation function.  As a function of $z_c=x+iv_ct$
and $z_s=x+iv_st$ (where $v_c$ and $v_s$ denote the Fermi velocities of the
charge and spin sector, respectively), the asymptotic form of $G_{nn}(x,t)=
\langle N(x,t) N(0,0) \rangle$ is
\begin{eqnarray}
   &&G_{nn}(x,t)= 
        n_e^2  
        + \frac{B_\downarrow \cos(2k_{F,\downarrow}x+\delta_\downarrow)}
             {|z_c|^{2\gamma_{\downarrow,c}} |z_s|^{2\gamma_{\downarrow,s}}}  
              \nonumber \\
        &+&  \frac{B_\uparrow \cos(2k_{F,\uparrow}x+\delta_\uparrow)}
             {|z_c|^ {2\gamma_{\uparrow,c}} |z_s|^ {2\gamma_{\uparrow,s}}}
         +  \frac{B_n \cos(2(k_{F,\uparrow}+k_{F,\downarrow})x+\delta_n)}
             {|z_c|^{2\gamma_{n,c}} |z_s|^{2\gamma_{n,s}}}
         \nonumber , \\
\label{ddper}
\end{eqnarray}
with $k_{F,\uparrow,\downarrow}=\pi n_{\uparrow,\downarrow}$.  The
exponents are displayed in Table\ \ref{tab:gamma}.  Eq.~(\ref{ddper}) shows
the oscillating terms which are the most relevant ones asymptotically.  For
vanishing magnetization the momenta $k_{F,\downarrow}$ and $k_{F,\uparrow}$
coincide, and one has to consider logarithmic corrections in $x$ (see Ref.
\onlinecite{gischu:89}) -- this case will not be considered below.

Following Cardy \cite{card:84} one has to replace $|z_c|^2 \rightarrow x$
and $|z_s|^2 \rightarrow x$ to obtain $\langle N(x) \rangle_o$ from Eq.\ 
(\ref{ddper}).  The final result is
\begin{eqnarray}
\langle N(x) \rangle=
n_e &+& A_\downarrow \frac{\cos(2k_{F,\downarrow}x+\varphi_\downarrow)}
             {x^{\gamma_{\downarrow,c}+\gamma_{\downarrow,s}}}
        + A_\uparrow \frac{\cos(2k_{F,\uparrow}x+\varphi_\uparrow)}
             {x^ {\gamma_{\uparrow,c}+\gamma_{\uparrow,s}}}
        \nonumber \\
        &+& A_n \frac{\cos(2(k_{F,\uparrow}+k_{F,\downarrow})x+\varphi_n)}
             {x^{\gamma_{n,c}+\gamma_{n,s}}} .
\label{dopen}
\end{eqnarray}  
The correlation function $G_{\sigma \sigma}^z \sim \langle M(x,t)M(0,0)
\rangle$ with magnetization $M=N_\uparrow-N_\downarrow$ has the same
critical behaviour as $G_{nn}(x,t)$.  Therefore $\langle M(x) \rangle$ has
the same form as $\langle N(x) \rangle$, but with different coefficients
$A_\alpha$.


\subsection{Density Matrix Renormalization Group}
\label{sec:DMRG}

The density matrix renormalization group method (DMRG)
\cite{whit.92,whit.93} has become one of the most powerful numerical
methods for calculating the low-energy properties of one-dimensional
strongly interacting quantum systems.  The expectation values of equal-time
operators in the ground state, such as the local density or magnetization
which interest us here, can be calculated with very good accuracy on quite
large systems (on lattices of up to $L=400$ sites in this paper).  As we
will see, access to such large system sizes is essential for the real space
fitting method used to extract the coefficients and exponents of the
Friedel oscillations (see Sec.\ \ref{sec:fit_meth}).  In the DMRG, open
boundaries are also the most favourable type of boundary conditions
numerically: for a given number of states kept (which corresponds to the
amount of computer time needed) the accuracy in calculated quantitites such
as the ground state energy is, in general, orders of magnitude better for
open boundary conditions than for periodic boundary conditions.
\cite{whit.93,skne.97}

In this work the {\sl finite system} DMRG method is used: after the system
is built up to a given size using a variation of the {\sl infinite system}
method, a number of finite-system iterations are performed in which the
overall size of the system (i.e.\ the superblock) is kept fixed, but part
of the system (the system block) is built up.  Optimal convergence is
attained by increasing the number of states kept on each iteration, and the
convergence of the exact diagonalization step is improved by keeping track
of the basis transformations and using them to construct a good initial
guess for the wavevector.\cite{whit.96} For all calculations shown in this
paper, we have performed 5 iterations with a maximum of $m=600$ states
kept.  The resulting discarded weight of the density matrix was ${\cal
  O}(10^{-7})$ and below.

\begin{figure}
\vbox{%
  \centerline{\epsfig{file=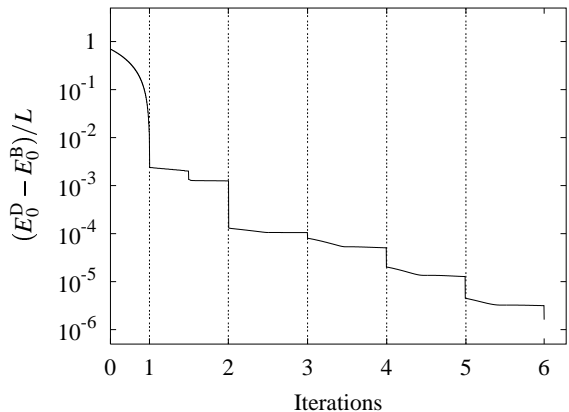,width=8cm}} \narrowtext
\caption{\label{fig:delta_egz}
  The difference between the ground state energy per site calculated with
  the DMRG, $E_0^{D}$, and the exact BA-energy, $E_0^{B}$, as a function of
  the number of DMRG iterations for a typical system ($L=400$, $n_e=0.55$,
  $n_\uparrow=0.35$, $U=10$ and $p=0$).}}
\end{figure}

\begin{figure}
\vbox{%
  \centerline{\epsfig{file=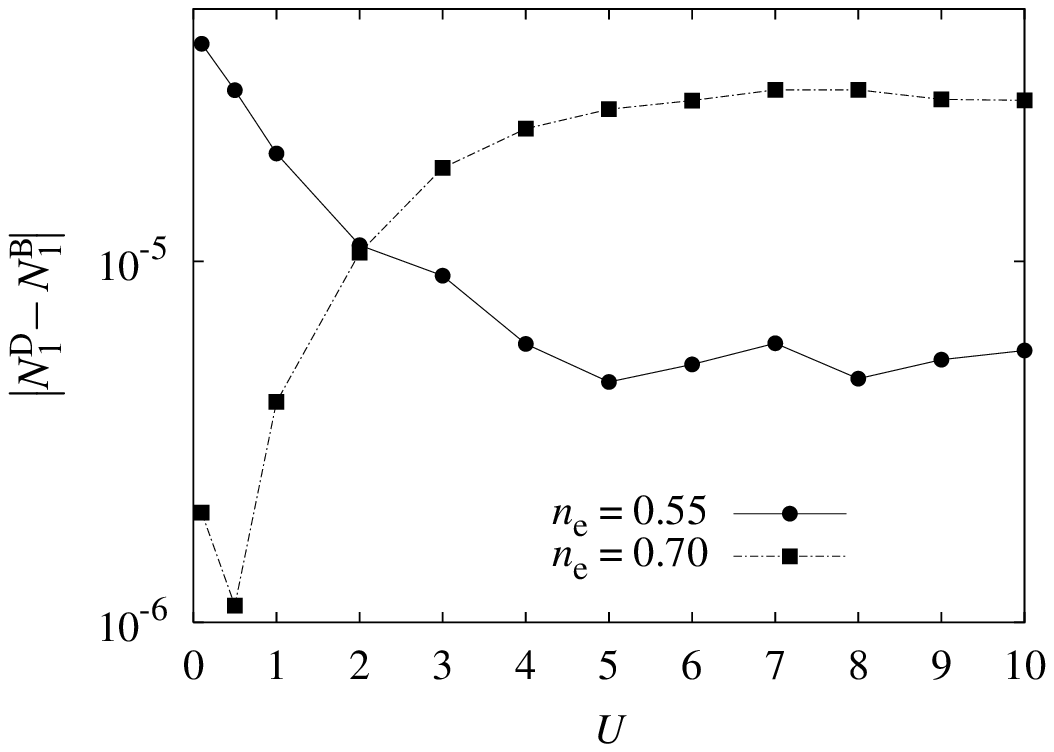,width=8cm}}
\caption{\label{fig:delta_n1}
  The difference between the expectation value of the density at site 1
  calculated with DMRG, $N_1^{D}$, and with BA, $N_1^{B}$, for the fillings
  $n_e=0.55$ and $n_e=0.70$ without boundary fields.}}
\end{figure}

\begin{figure}
\vbox{%
  \centerline{\epsfig{file=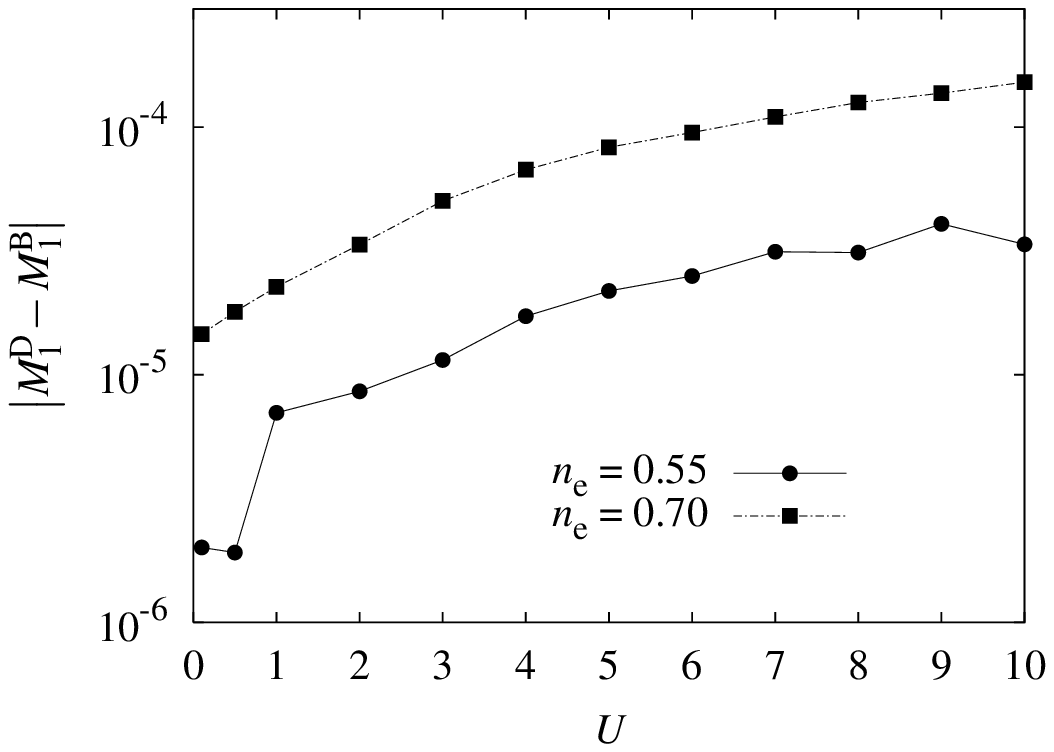,width=8cm}}
\caption{\label{fig:delta_m1}
  The difference between the expectation value of the magnetization at site
  1 calculated with DMRG, $M_1^{D}$, and with BA, $M_1^{B}$, for $n_e=0.55$
  and $n_e=0.70$, without boundary fields.}}
\end{figure}

We illustrate the convergence of the algorithm explicitly in Figs.
\ref{fig:delta_egz}--\ref{fig:delta_m1}.  One finds that for all the
parameters which are used in this paper the ground-state energy per site is
accurate to ${\cal O}(10^{-6})$ or less, Fig.\ \ref{fig:delta_egz}, while
the expectation value of the density at the first site (or at the last
site, due to symmetry) is accurate to ${\cal O}(10^{-5})$, as can be seen
in Fig.\ \ref{fig:delta_n1}.
\narrowtext
\begin{table*}
\caption{Exponents $\gamma_{\alpha,c}$ and $\gamma_{\alpha,s}$ as 
a function of the elements of the dressed charge matrix.
\label{tab:gamma}}
\renewcommand{\arraystretch}{1.2}
\begin{tabular}{ccc}
 &  $\gamma_{\alpha,c}$ & $\gamma_{\alpha,s}$ \\
 \tableline
 $\alpha=\downarrow$  &  $(Z_{cc}-Z_{sc})^2$  &  $(Z_{cs}-Z_{ss})^2$  \\
 $\alpha=\uparrow$ & $Z_{sc}^2$ & $Z_{ss}^2$ \\
 $\alpha=n$ & $Z_{cc}^2$  & $Z_{cs}^2$ \\ 
\end{tabular}
\end{table*}
The magnetization shows an analogous behaviour but with results correct up
to ${\cal O}(10^{-4})$, Fig.\ \ref{fig:delta_m1}.  It is also interesting
to note that for $U\le 10$ there is no strong $U$-dependence of the
accuracy of either the density or the magnetization expectation values.

We now want to examine the effect of switching on the boundary fields
simultaneously at the first and last sites on the quality of the DMRG
results.  In order to do this, we compare the mean density $\langle N_1(p)
\rangle$ from DMRG calculations with Bethe Ansatz results calculated in the
thermodynamic limit. Within the Bethe Ansatz, the mean density is
calculated from the derivative of $\langle {\cal H} \rangle$ with respect
to the boundary field $p$.
\begin{figure}
\vbox{%
  \centerline{\epsfig{file=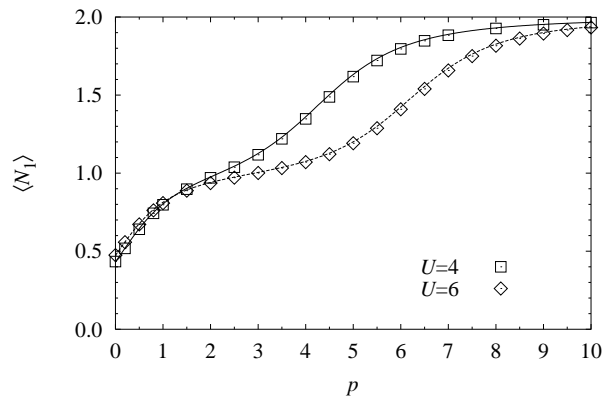,width=8cm}}
\caption{\label{fig:n1ofp}
  The expectation value of $\langle N_1\rangle$ for different values of the
  interaction $U$ and boundary fields $p$ on a $L=100$ lattice with
  $n_e=0.55$ and $n_\uparrow=0.35$.  The solid lines represent the exact
  solutions in the thermodynamic limit, while the symbols are DMRG
  results.}}
\end{figure}
The numerical results, shown in Fig.\ \ref{fig:n1ofp} on an $L=100$ lattice
for electron density $n_e=0.55$, $n_\uparrow=0.35$ and for two values of
the interaction $U$, are again in very good agreement with the values for
the thermodynamic limit.  The difference $N_1^{\rm D}-N_1^{\rm B}$ is now
${\cal O}(10^{-3})$.  While this seems to be worse than the $p=0$ case, we
have neglected finite-size corrections to $\langle N_1 \rangle$ since we
have compared to thermodynamic limit Bethe Ansatz calculations.  If we
explicitly take the $1/L$ corrections to the Bethe Ansatz values into
account, we find agreement to ${\cal O}(10^{-5})$.  This value is already
smaller than the $1/L^2$-corrections so that finally we state that $\langle
N_1 \rangle$ is accurate to ${\cal O}(10^{-4})$.


\section{Friedel Oscillations}
\label{sec:fr_osc}

In general, the presence of an impurity or boundary in a one-dimensional
fermion system leads to Friedel oscillations in the density, which have the
general form
\begin{equation}
\delta \rho(x) \sim {\cos(2k_F x+\varphi) \over x^\gamma} \ ,
\label{eq:drho}
\end{equation}
where the exponent $\gamma$ depends on the interaction.  In addition to
numerical studies of these oscillations for spinless fermions
\cite{schmeck:96} and Kondo--Systems \cite{shueni:96}, several theoretical
attempts have been made to clarify the role of interaction.
Using bosonization it is possible to obtain the asymptotic exponents as a
function of the interaction parameters and corrections to the power--law
behaviour of Eq.\ (\ref{eq:drho}).\cite{fago:95,eggra:95} CFT results were
used to calculate the interaction dependence of the exponent $\gamma$ for
interacting spinless fermions. \cite{wavp:96}
Here we start with noninteracting fermions to obtain some conjectures for
the connection between the explicit form of the Friedel oscillations and
Bethe Ansatz results.  These conjectures will then be checked using the
DMRG results.


\subsection{Noninteracting Fermions}

By considering only spin-$\uparrow$ electrons without any boundary
potential, one can easily obtain the expectation value of the electron
density:
\begin{equation}
N(x)={N_\uparrow+{1 \over 2} \over L+1}-{1 \over 2 (L+1)}
\frac{\sin \left(2 \pi x {N_\uparrow+1/2 \over L+1} \right)}
{\sin\left({\pi x \over L+1}\right)}.
\end{equation}
In the limit $L \to \infty$ and $x \ll L$ the density is given by
\begin{eqnarray}
N(x) &=&n_\uparrow-{n_\uparrow-{1 \over 2} \over L}-{\sin \left(2 \pi x 
 (n_\uparrow-{n_\uparrow-{1 \over 2} \over L}) \right) \over 2 \pi x}
 \nonumber \\ 
 &=& n_\uparrow-{\theta^c_0 \over L}-{\sin \left(2 \pi x 
 (n_\uparrow-{\theta^c_0 \over L}) \right) \over 2 \pi x} ,
\label{free}
\end{eqnarray}
with $\theta^c_0$ defined in Eq.\ (\ref{eq:tpc1}).  The density $N(x)$ can
also be calculated explicitly when the boundary field $p=1$.  It then has
the same structure as Eq.\ (\ref{free}).

If one assumes that the Friedel oscillations in the interacting system have
an analogous structure to those in the non-interacting system, one can
combine Eqs.\ (\ref{dopen}) and (\ref{free}) to obtain the following
conjectures for the finite-size shifts of the average density, average
magnetization and the characteristic wave vectors in the {\sl interacting}
system:
\begin{equation}
\overline{n}=n_e-{\theta_n \over L}, \qquad
\overline{m}=m-{\theta_m \over L}
\label{eq:nmc}
\end{equation}
\begin{equation}  
k_{F,\downarrow}=\pi \left(n_\downarrow-{\theta_\downarrow \over L}\right), 
\qquad
k_{F,\uparrow}=\pi \left(n_\uparrow-{\theta_\uparrow \over L}\right)
\label{eq:kc}
\end{equation}
with $\theta_\downarrow=\theta^s_p, \theta_\uparrow= \theta^c_p-\theta^s_p,
\theta_n=\theta^c_p$ and $\theta_m= \theta^c_p-2\theta^s_p$.


\subsection{Fit procedure}
\label{sec:fit_meth}

Previously, several methods have been used to obtain asymptotic exponents
of correlation functions using numerical
data.\cite{sopa:90,qili:95,ogasushi:91} All of these methods use the
$L$--dependence of the fourier-transformed correlation functions near the
relevant peaks $k_\alpha$ $(\alpha=\uparrow,\downarrow,n)$ in Fourier space
to extract the exponents.  Due to the fact that only systems with periodic
boundary conditions were considered, the $k_\alpha$ were all independent of
$L$.  This $L$-independence seems to be crucial for these methods to work;
we were not able to extract a reasonable exponent with any of these methods
on a system with open boundary conditions.

Therefore, we fit the DMRG results for $N(x)$ and $M(x)$ to the real-space
test function
\begin{eqnarray}
  \label{eq:fitfunc}
  f(x)=\left\{ \begin{array}{c} \overline{n} \\
        \overline{m} \end{array} \right\}
      +\sum_{\alpha \in \{\uparrow,\downarrow,n\}} 
  {A_\alpha \sin(2 k_\alpha x +\varphi_\alpha) \over x^{\gamma_\alpha}}
  \nonumber \\
  +{A_\alpha \sin(2 k_\alpha x +\varphi_\alpha) \over (L+1-x)^{\gamma_\alpha}} 
\end{eqnarray}
which explicitly includes the momenta as fit parameters.  Here the second
term is included due to symmetry.  There are a total of 13 fit parameters
in this function, a prohibitively large number to do a simultaneous fit of
all parameters.  However, if we only consider systems in which the three
peaks in the Fourier--spectrum are well seperated, there is an effective
fit of 4 parameters to every peak.  As we will see, the peak at $k_n$ is
supressed for small $U$, reducing the number of fit parameters to 9 for
only two peaks.  The amplitudes $A_\alpha$ will be assumed to be positive,
with any sign given by the phase $\varphi_\alpha$.  We fit to the
magnetization $M(x)$ and the density $N(x)$ independently.

The right side of Eq.\ (\ref{free}) is only valid for $x \ll L$.  In
addition, the CFT results are valid only asymptotically for large
distances.  As a consequence and compromise, we do not use the density
information on the first five and last five lattice points.

We perform the least squares fit in two stages.  In the first stage, the
start parameters of the subsequent fit are determined using simulated
annealing.  The final fit is performed using a combined Gauss--Newton and
modified Newton algorithm (using the NAG routine E04FCF).  To estimate the
fit error, 10 fits are performed for each system with 10\% of the points
randomly excluded from each fit.


\subsection{Results}
\label{sec:results}

Before discussing the results for the Friedel oscillations in detail, we
make some general comments on the numerical results.  As described in Sec.\ 
\ref{sec:fit_meth}, we calculate the quantities for the density and
magnetization oscillations by applying a 13-parameter fit.  Fitting to this
many parameters requires the use of large system sizes.  While the
numerical expense for the DMRG procedure grows linearly with the system
size for a fixed number of states kept, the accuracy in the energy and in
the local density and magnetization decreases with the system size,
especially in the Luttinger liquid regime. \cite{skne.97} We have compared
the accuracy of the DMRG results with the accuracy and convergence of the
fitting procedure for different lattice sizes from $L=300$ to $L=500$ and
have decided that $L=400$ yields optimal results for the amount of
computing power available.  However, results within this range of system
sizes are in agreement to within DMRG and fitting errors.

Another important issue is the influence of the boundary potentials on the
fitting method and on the Friedel oscillations (discussed in Sec.\ 
\ref{sec:ne55p}).  As the boundary potential $p$ is increased, bound states
will develop at site 1 and site $L$.\cite{befr:97} In order to avoid these
bound states in the fitting procedure, one has to enlarge the range in
which the local density $N(x)$ is disregarded from 5 (i.e. $x, L-x-1 \le
5$) at $p=0$ to about 20 at $p=9.9$.

The discussion of the next three sections will focus on comparing the
BA/CFT predictions for the different fit parameters with the DMRG results,
especially on checking the conjectures from Eqs.\ (\ref{eq:nmc}) and
(\ref{eq:kc}).  We also compare the numerical results to the different
exactly known values for different limits such as the limit of
noninteracting fermions, $U\to 0$.


\subsubsection{Density $n_e=0.55$}
\label{sec:ne055}

\begin{figure}
  \epsfxsize=0.9\textwidth \epsfbox{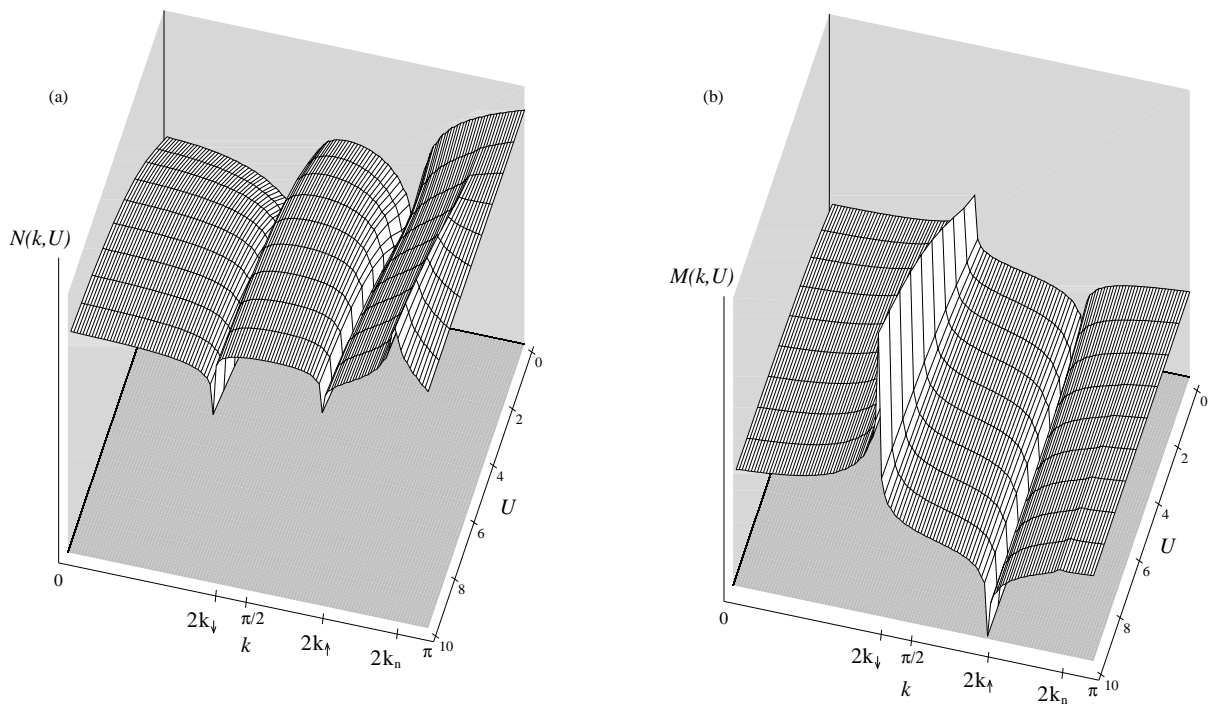}
\caption{Fourier transformation of (a) the density $N(x)$ and (b) the
  magnetization $M(x)$ for several values of $U$ at the density $n_e=0.55$
  and $n_\uparrow=0.35$. (The $k=0$ values are excluded.)
\label{fig:ftne55}}
\end{figure}

We first examine the Fourier transform of the local electron density
$N(x)$, defined as
\begin{equation}
N(k,U)=\sum_{x=1}^L \cos\left(k(x-{1 \over 2})\right) N(x,U)
\label{eq:ft}
\end{equation}
with $k={ 2 \pi j \over L}$ and $j=0,\ldots,{L \over 2}-1$.  (Due to
symmetry $N(k,U)$ vanishes for odd multiples of ${\pi \over L}$.)  The
quantity $N(k,U)$ is displayed in a three-dimensional plot in Fig.\ 
\ref{fig:ftne55}(a).  Distinct peaks at the three wave vectors,
$k_\uparrow$, $k_\downarrow$, and $k_n$ can clearly be seen.  Note that we
have chosen $n_e$ and $n_\uparrow$ so that these three peaks are
well-separated.  However, the peak at $k_n$ becomes very lightly weighted
and therefore ill-defined for $U<1$.  In fact, we have found that it is not
possible to locate the third momentum $k_n$ for $U<1$ using the
13-parameter fit procedure described in Sec.\ \ref{sec:fit_meth}.
Therefore, we fit $N(x)$ using only 9 parameters for $U=0.1$ and $U=0.5$.
We display the fourier-transformed magnetization $M(k,U)$, defined
analogously, in Fig.~\ref{fig:ftne55}(b).  Here the $k_n$ peak is even more
\end{multicols}
\widetext
\begin{multicols}{2}
\narrowtext
\begin{figure}
\vbox{%
  \centerline{\epsfig{file=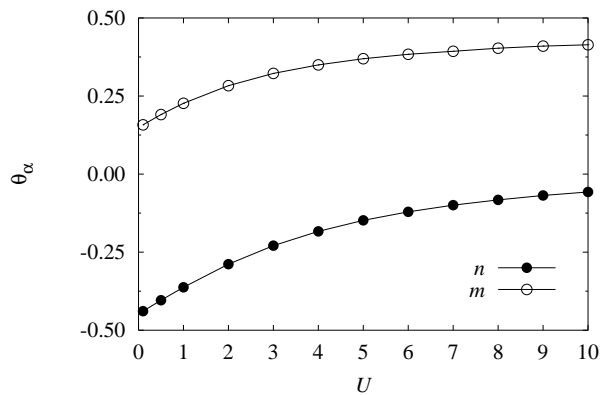,width=8cm}}
\caption{
  Difference $\theta_\alpha=(\alpha-\overline{\alpha})L$ for the density
  ($\alpha=n$) and the magnetization ($\alpha=m$) as a function of $U$ for
  $n_e=0.55$.  The solid lines are the Bethe Ansatz conjectures
  $\theta_\alpha$.
\label{fig:mean55}
}}
\end{figure}
\noindent%
poorly defined, and, in fact, is at best barely discernible, even at large
$U$.  Therefore, it is only possible to fit to two peaks using 9 parameters
in the entire region from $U=0$ to $U=10$.  For these fit procedures, we
have found that the mean-squared deviation $\sigma^2 =
\sum_{x}(N(x)-f(x))^2$ is between ${\cal O} (10^{-7})$ and ${\cal
  O}(10^{-6})$ for the density and between ${\cal O}(10^{-6})$ and ${\cal
  O}(10^{-5})$ for the magnetization for all $U$ values.  These limits will
apply to all the fit results shown in this paper.

In Fig.\ \ref{fig:mean55} we show the $1/L$ corrections of the mean values
$\overline{n}$ and $\overline{m}$ calculated with the DMRG and from Bethe
Ansatz using the conjectures in Eq.\ (\ref{eq:nmc}).  One can see that
there is quite good agreement between the two calculations for all $U$.

The comparison of the exact asymptotic exponents at the different momenta
with the numerical results is one of the most interesting and important
features of this work because similar methods will then be able to be used
to calculate properties not directly predictable with CFT/BA, and to treat
systems that are not Bethe Ansatz solvable.

\begin{figure}
\vbox{%
  \centerline{\epsfig{file=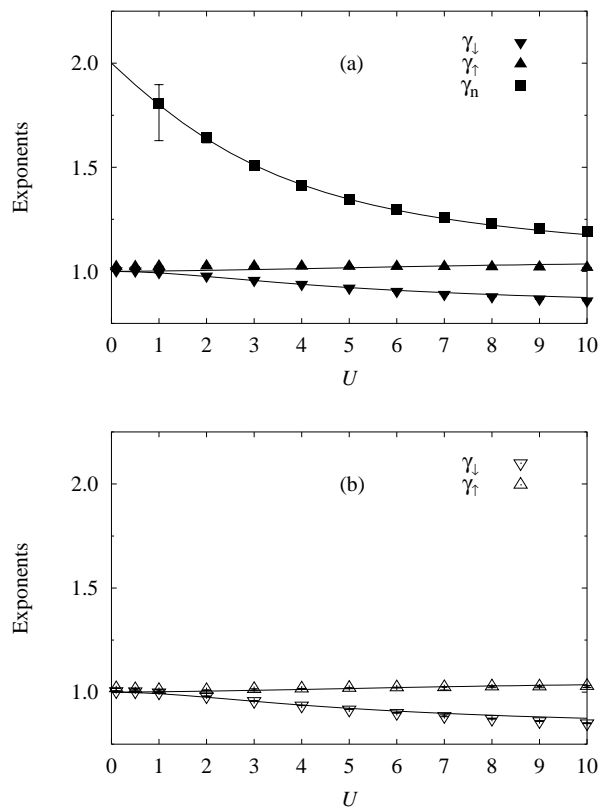,width=8cm}}
\caption{
  Exponents for the three peaks as a function of $U$ fitted for (a) the
  particle density and (b) the two peaks of the magnetization for
  $n_e=0.55$.  The solid lines denote the CFT predictions from the Bethe
  Ansatz solution.
\label{fig:exp55}
}}
\end{figure}

The exponents at $k_\downarrow$ and $k_\uparrow$, extracted from the
density as well as from the magnetization data, are shown in
Fig.~\ref{fig:exp55}.  The difference between the fit--exponents and the
CFT-prediction is less than 2\% for the fits to the density and less than
3\% for the fits to the magnetization.  As mentioned above, we can obtain
an exponent for the peak at $k_n$ from the density fit for $U \ge 1$ only,
due to its small weight especially for small $U$.  At $U=1$ the large error
bars reflect a poor fit.  Here we have only considered $U \le 10$, for
which $\gamma_n>\gamma_\uparrow$.  A further increase in $U$ would lead to
a region where $\gamma_n<\gamma_\uparrow$.  A crossover between these two
regions will be seen for the $n_e=0.70$ in the next section, for which it
occurs at a somewhat lower $U$.
Due to the fact that we obtain only three independent exponents from the
fitting procedures, it is not possible to determine all of the elements of
the dressed charge matrix ${\bf Z}$.  In fact, only the following
combinations are relevant for the three exponents extracted:
$Z_{cc}^2+Z_{cs}^2,Z_{ss}^2+Z_{sc}^2$ and $Z_{cc}Z_{sc}+Z_{cs}Z_{ss}$.
\cite{ogasushi:91} It would be possible to uniquely determine all of the
independent elements of the dressed charge matrix with additional
information from, for example, any susceptibility \cite{frko:90} or from
another correlation function with a different set of critical exponents.
Relationships between the elements of the dressed charge matrix ${\bf Z}$
and the parameters of the Tomonaga--Luttinger model are given in Ref.
\onlinecite{peso:93}.

Within the framework of CFT, the amplitudes $A_\alpha$ are completely
undetermined.  However, the form--factor approach \cite{less:97} may lead
to explicit results for the amplitudes in the future.  For example, a
conjecture of Lukyanov and Zamolodchikov \cite{lukyza:97} concerning the
amplitude of the spin--spin correlation functions of the XXZ chain was
recently confirmed by a fit to DMRG results. \cite{hifu:98}

\begin{figure}
\vbox{%
  \centerline{\epsfig{file=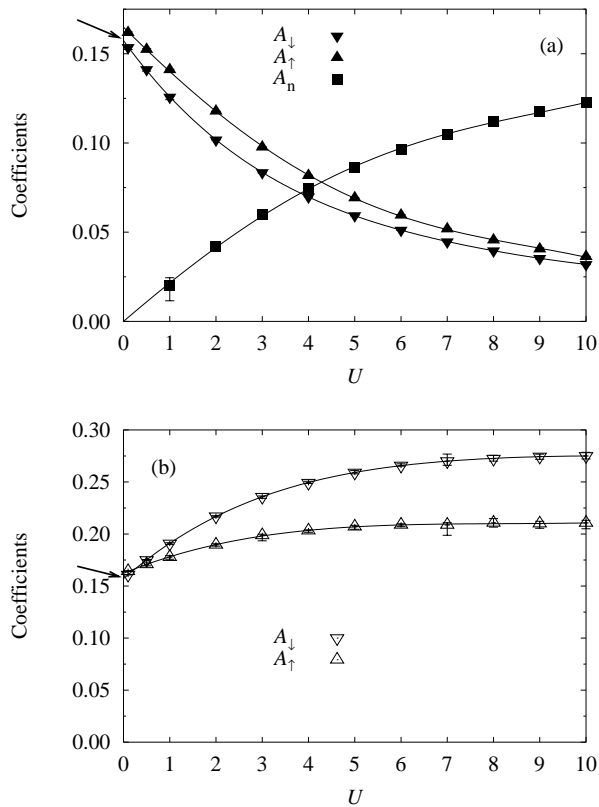,width=8cm}}
\caption{
  Amplitudes $A_\alpha$ fitted to (a) the density and (b) the magnetization
  as a function of $U$ for $n_e=0.55$.  The solid lines are guides to the
  eye.  The arrow denotes the value for the noninteracting fermions (=$1
  \over 2 \pi$).  Note that the amplitudes are, in general, different for
  the density and the magnetization.
\label{fig:koeff55}
}}
\end{figure}

At this point, however, the fit results can only be compared to
noninteracting fermions ($U=0$), for which $A_\downarrow=A_\uparrow={1
  \over 2 \pi}$.  As can be seen in Fig.~\ref{fig:koeff55}, this value is
in relatively good agreement (4\% deviation) with the fit results.  In
addition, $A_n=0$ for $U=0$, in agreement
\noindent%
with the $U=0$ extrapolated value in Fig.\ \ref{fig:koeff55}(a).  The large
error bars in $A_n$ at $U=1$ are due to the difficulty in fitting the $k_n$
peak for small $U$.  Since the fitting procedure seems to work well for the
exponents, and the amplitudes yield the correct $U=0$ limit, we feel that
the calculation of the amplitudes is under good control.  This is therefore
the first determination of the qualitative as well as quantitive behaviour
of these amplitudes.

\begin{figure}
\vbox{%
  \centerline{\epsfig{file=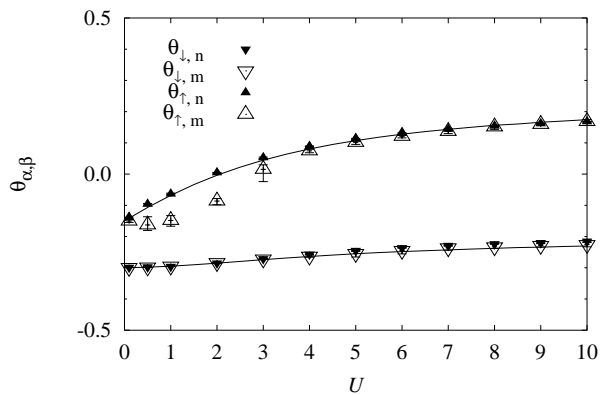,width=8cm}}
\caption{
  Difference $\theta_{\alpha,\beta}=(n_\alpha-{k_{\alpha,\beta} \over
    \pi})L$ for $\alpha=\downarrow,\uparrow$ as a function of $U$ for the
  density $n$ and magnetization $m$ for $n_e=0.55$.  The solid lines denote
  the Bethe Ansatz conjectures $\theta_\alpha$ [see Eq.\ (\ref{eq:kc})].
\label{fig:mom55}
}}
\end{figure}

The exact position of the momenta $k_\uparrow$ and $k_\downarrow$ are
further fit parameters.  The $1/L$ corrections to the {\it thermodynamic
  value} are plotted in Fig.~\ref{fig:mom55}.  The fit values agree well
with the Bethe Ansatz conjectures, except for the correction to
$k_{\uparrow,m}$, which deviates from the Bethe Ansatz value for $U<4$.
These deviations are probably due to problems with the fit.  Note also that
the Bethe Ansatz results for $\theta_\alpha$ are correct only up to ${\cal
  O}(1/L)$.  The momentum $k_n$, which is not shown, is another independent
fit parameter.  The fit error in $k_n$ extracted from the fit to the
density $N(x)$ is rather large for $U<4$. This is due to the fact that the
peak at $k_n$ is not well--defined enough in this region to obtain the
$1/L$ corrections to this momentum. Nevertheless the agreement between the
fit values and the Bethe Ansatz conjectures is very good for%
\noindent
$U>4$.


\subsubsection{Density $n_e=0.70$}
\label{sec:ne070}

In this section, we examine the same quantities as in the previous section
at a density of $n_e=0.70$.  A treatment of this density is interesting for
a number of reasons.  Since we use the same numerical parameters for both
densities, we can examine the density dependence of the error in truncating
the Hilbert space using the DMRG.  This density is also interesting because
the BA/CFT calculations predict that the crossover between the $\gamma_n$
and $\gamma_\uparrow$ exponents will take place within the range of
interaction treated here, $U=0\ldots 10$.

\begin{figure}[ht]
  \epsfxsize=0.9\textwidth \epsfbox{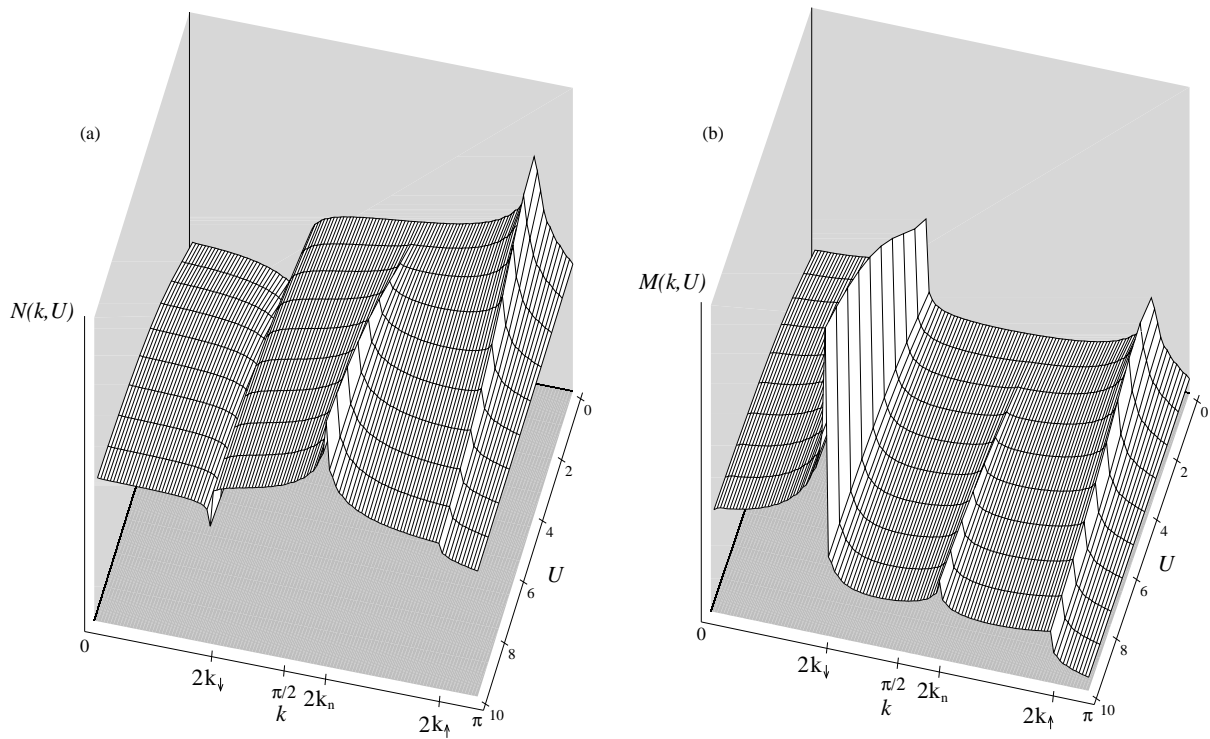}
\caption{Fourier transformation of (a) the density $N(x)$ and (b) the
  magnetization $M(x)$ for several values of $U$ for the density $n_e=0.70$
  with $n_\uparrow=0.55$ ($k=0$--values excluded).  Note that the momentum
  $k_n$ is located between $k_\downarrow$ and $k_\uparrow$.
\label{fig:ftne70}}
\end{figure}

The Fourier transforms of the local density $N(x)$ and the magnetization
$M(x)$ are shown in Fig.~\ref{fig:ftne70}.  Note that the momentum $k_n$
(wrapped back to the range $0$ to $\pi$) is now located between
$k_\downarrow$ and $k_\uparrow$.  As can be seen in
Fig.~\ref{fig:ftne70}(a), the peak in $N(x)$ at $k_n$ is not well-defined
for $U<1$, so the $U=0.1$ and $U=0.5$ data are fitted using 9 parameters to
fit two peaks.  However, the $k_n$ peak in $M(x)$ is now more well-defined
than for $n_e=0.55$, as can be seen in Fig.~\ref{fig:ftne70}(b), and it is
now possible to fit to all three momenta for $U \ge 2$.  For smaller values
of $U$ ($U=0.1$, $U=0.5$ and $U=1$), a nine-parameter fit is again made to
two peaks.

\begin{figure}
\vbox{%
  \centerline{\epsfig{file=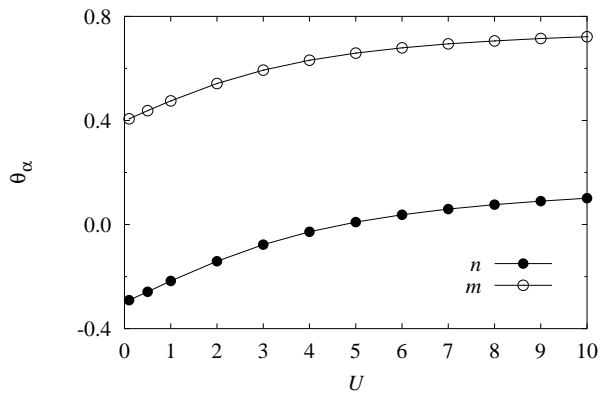,width=8cm}}
\caption{
  Difference $\theta_\alpha=(\alpha-\overline{\alpha})L$ for density
  ($\alpha=n$) and magnetization ($\alpha=m$) as a function of $U$ for
  $n_e=0.7$.  The solid lines are the Bethe Ansatz conjectures
  $\theta_\alpha$.
\label{fig:mean70}
}}
\end{figure}

\begin{figure}
\vbox{%
  \centerline{\epsfig{file=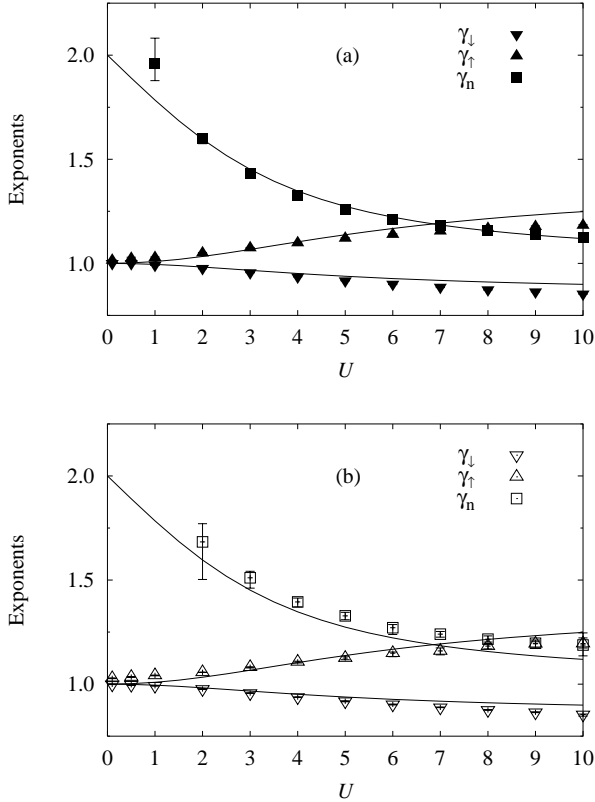,width=8cm}}
\caption{
  Exponents for the three peaks of (a) the particle density and (b) the
  magnetization as a function of $U$ for $n_e=0.7$.  The solid lines denote
  the CFT predictions from the Bethe Ansatz solution.
\label{fig:exp70}
}}
\end{figure}

The $1/L$ corrections to the mean values of the density and the
magnetization, shown in Fig.~\ref{fig:mean70}, are in very good agreement
with the BA conjectures, thereby providing a further confirmation of the
predictions of Eqs.\ (\ref{eq:nmc}) and (\ref{eq:kc}).  The exponents
extracted from the fit are shown in Fig.~\ref{fig:exp70}.  The expected
crossing of the two largest exponents at $U \approx 7$ can clearly be seen.
For $\gamma_\downarrow$ and $\gamma_\uparrow$ obtained from the density
fit, Fig.~\ref{fig:exp70}(a), the deviation from the CFT results is about
5\% at most, with the largest errors occuring for $U>6$, especially in
$\gamma_\uparrow$.  As can be seen in Fig.~\ref{fig:ftne70}(a), the peak at
$k_\uparrow$ in $N(k,U)$ gets weaker for
\end{multicols}
\widetext
\begin{multicols}{2}
  \narrowtext
\noindent%
larger $U$, leading to a less effective fit.  The agreement of the fitted
exponents $\gamma_n$ with the CFT predictions is much better, with a
deviation from the BA/CFT values of less than 1\% for $U>1$.  The exponents
obtained from fits to the magnetization, Fig.~\ref{fig:exp70}(b), show
deviations of up to about 6\% from the BA/CFT results.  For $U=9$ and
$U=10$, the deviation is largest and the exponents coincidentally take on
the same value.  Again, this is probably due to larger errors in the fit
because the peak at $k_\uparrow$ becomes weaker at larger $U$.  One can see
that the peaks at $k_\uparrow$ and $k_n$ are much less heavily weighted
than the peak at $k_\downarrow$ at large $U$.

\begin{figure}
\vbox{%
  \centerline{\epsfig{file=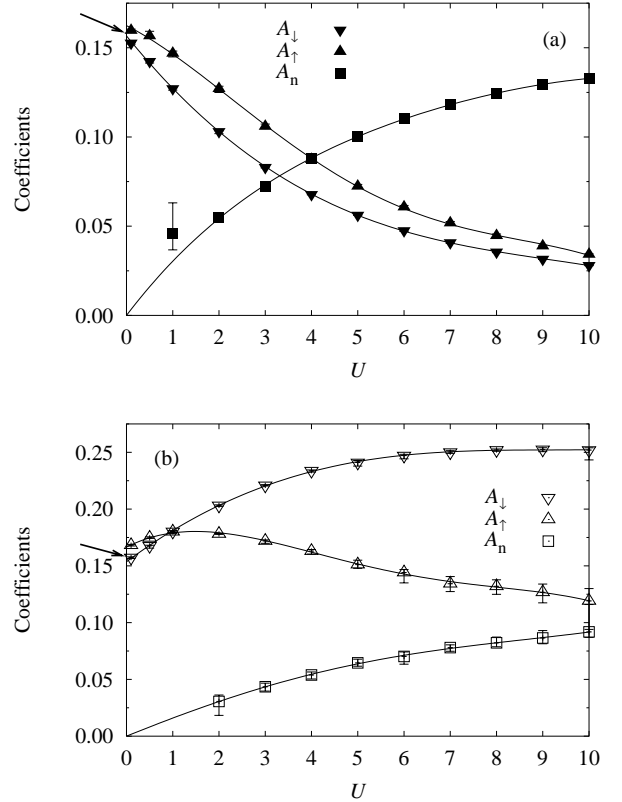,width=8cm}}
\caption{
  Amplitudes $A_\alpha$ fitted to (a) the density and (b) the magnetization
  as a function of $U$ for $n_e=0.7$.  The solid lines are a guide to the
  eye.  The arrow denotes the value $1\over 2 \pi$ of the noninteracting
  fermions.  Note that the amplitudes are, in general, different for the
  density and the magnetization.
\label{fig:koeff70}
}}
\end{figure}

The amplitudes $A_\downarrow$ and $A_\uparrow$ extracted from the fit to
the density, shown in Fig.~\ref{fig:koeff70}(a), decrease monotonically
with increasing $U$.  The $U=0$ values agree with the exactly known value
of $1/2\pi$ to within about 4\%.  The amplitude $A_n$, on the other hand,
increases with increasing $U$.  Its $U\rightarrow 0$ extrapolation agrees
well with the value zero of the noninteracting fermions if the $U=1$ point,
which cannot be very accurately determined, is excluded.  The behaviour and
even the quantitative values of all three coefficients are quite similar to
the $n_e=0.55$ case shown previously in Fig.~\ref{fig:koeff55}(a).  The
amplitudes obtained from the fit to the magnetization $M(x)$ are shown in
Fig.~\ref{fig:koeff70}(b).  The amplitude $A_\downarrow$ behaves similarly
to the $n_e=0.55$ case [Fig.~\ref{fig:koeff55}(b)] in that it increases
with increasing $U$, but $A_\uparrow$ shows different behaviour in that it
reaches a maximum at $U\approx 1$ and then decreases.  Both fit amplitudes
yield the $U=0$ value of $1/2\pi$ to within 6\%.  The amplitude $A_n$ for
the summed momenta, which could not be determined for $n_e=0.55$, increases
monotonically with $U$, and its $U\rightarrow 0$ extrapolation agrees well
with the value for noninteracting fermions, $A_n = 0$.

\begin{figure}
\vbox{%
  \centerline{\epsfig{file=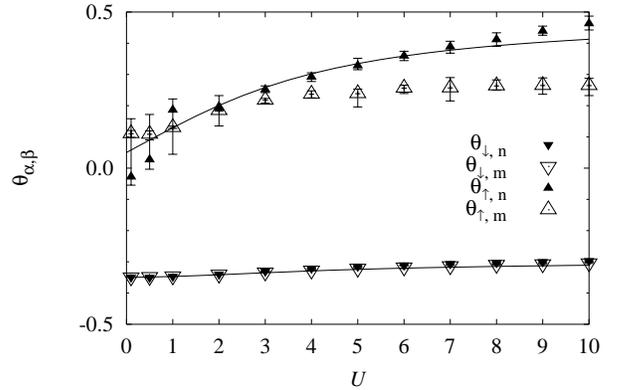,width=8cm}}
\caption{
  Difference $\theta_{\alpha,\beta}=(n_\alpha-{k_{\alpha,\beta} \over
    \pi})L$ for $\alpha=\downarrow,\uparrow$ as a function of $U$ for the
  density $n$ and magnetization $m$ for $n_e=0.7$.  The solid lines denote
  the Bethe Ansatz conjectures $\theta_\alpha$ [see Eq.\ (\ref{eq:kc})].
\label{fig:mom70}
}}
\end{figure}

The $1/L$ corrections to the momenta $k_\downarrow$ fitted for the density
and the magnetization, shown in Fig.~\ref{fig:mom70}, are in good agreement
with the Bethe Ansatz conjecture.  The agreement is also fairly good for
$\theta_{\uparrow,n}$, although the error of the fit is rather large for
small $U$. However, the $\theta_{\uparrow,m}$ fit results do not match well
with the conjecture.  As we have seen in Fig.~\ref{fig:ftne70}(b), the
$k_n$ and $k_\downarrow$ peaks in $M(x)$ have much lower amplitudes than
the $k_\uparrow$ peak, leading to lower accuracy in the fitting procedure.
The fit results%
\noindent%
for the exponents, Fig.~\ref{fig:exp70}(b) also had a rather large
deviation from the CFT results in this regime.  Thus it is not possible to
confirm or deny the conjecture concerning the shift of $k_\uparrow$ for
$M(x)$.

For $k_n$ the situation is even worse. Both density and magnetization fits
lead to large fit errors for $U<4$.  The deviation from the Bethe Ansatz
conjectures is about $0.08$ in both fits, outside the range of the $1/L$
correction to $\theta_n$.

In summary, for $n_e=0.70$ the DMRG results for the mean density
(magnetization, respectively) and the exponents are in good agreement with
exact results from BA/CFT, further confirming Conformal Field Theory.

A detailed examination of the convergence of the DMRG shows that the
numerical accuracy is actually slightly worse than the $n_e=0.55$ case, but
this could be improved by increasing the number of states kept in the DMRG.
We therefore expect to be able to apply these techniques reliably to obtain
the boundary exponents and coeffients of other, non-Bethe-Ansatz solvable
one-dimensional models.


\subsubsection{Effect of boundary potentials}
\label{sec:ne55p}

We now examine the effect of the boundary potential $p$ on the Bethe Ansatz
predictions.  We set $n_e=0.55$, $n_\uparrow=0.35$, and $U=8$ and consider
six $p$ values, for which four qualitatively different Bethe Ansatz
solutions exist.  For $p=-0.5,\, 0,\, 0.5$, no bound state is present in
the Bethe Ansatz ground state; we examine both repulsive and attractive
fields.  The values $p=2.6$ and $p=6$ are in a region in which the Bethe
Ansatz has two complex $k$ solutions, one for each boundary potential at
site $x=1$ and $x=L$.  The $p=6$ ground state configuration contains, in
addition, two complex $\lambda$ solutions.  Finally, for $p=9.9$ there are
four complex $k$ and two complex $\lambda$ solutions, corresponding to a
bound pair of electrons at each end of the chain.  Details of the structure
of the ground state as a function of $p$ are given in Ref.\ 
\onlinecite{befr:97}.

The $p$-dependence of $N(k,p)$ and $M(k,p)$ at $U=8$ is not as strong as
the $U$-dependence of $N(k,U)$ and $M(k,U)$ found previously.  Since we
have chosen a fairly large $U$, the peaks in the density $N(x)$ have enough
weight to fit all three momenta.
As was the case for $p=0$, the peak at $k_n$ in the magnetization is not
pronounced enough to be fitted at any $p$.

\begin{figure}
\vbox{%
  \centerline{\epsfig{file=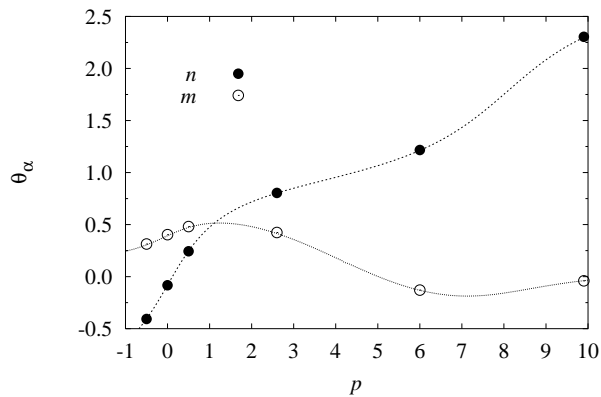,width=8cm}}
\caption{
  Difference $\theta_\alpha=(\alpha-\overline{\alpha})L$ for density
  ($\alpha=n$) and magnetization ($\alpha=m$) as a function of $p$ for
  $U=8$, $n_e=0.55$ and $n_\uparrow=0.35$.  The solid lines are the Bethe
  Ansatz conjectures for $\theta_\alpha$.
\label{fig:meanp55}
}}
\end{figure}

The $1/L$ corrections to the mean values of the density and magnetization,
Fig.~\ref{fig:meanp55}, are again in very good agreement with the Bethe
Ansatz conjectures, showing that Eqs.\ (\ref{eq:nmc}) and (\ref{eq:kc}) are
valid even in the different physical regions described above.  Within the
BA/CFT calculations, the values of the exponents are independent of the
boundary potentials $p$.  This agrees with the DMRG results, which we do
not show here: the range of the exponents varies by at most 2\% from the
exact values for $p=0$ (after a larger number of lattice points are
discarded from the fits in order to avoid the bound states at the ends).

The amplitudes also have no significant $p$-dependence.  The density fit
yields $A_\downarrow \approx A_\uparrow \approx 0.045$ and $A_n \approx
0.12$,%
\begin{figure}
\vbox{%
  \centerline{\epsfig{file=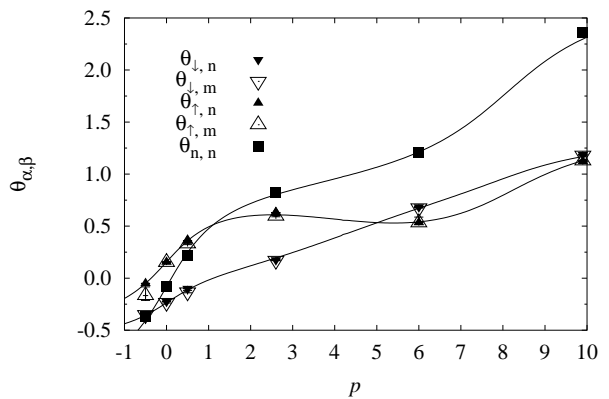,width=8cm}}
\caption{
  Difference $\theta_{\alpha,\beta}=(n_\alpha-{k_{\alpha,\beta} \over
    \pi})L$ for $\alpha=\downarrow,\uparrow,n$ as a function of $p$ for the
  density $n$ and magnetization $m$ for $U=8$, $n_e=0.55$ and
  $n_\uparrow=0.35$.  The solid lines denote the Bethe Ansatz conjectures
  $\theta_\alpha$ [see Eq.\ (\ref{eq:kc})].
\label{fig:mom55p}
}}
\end{figure}
\noindent%
while the magnetization fit yields $A_\downarrow \approx 0.28$ and
$A_\uparrow\approx 0.21$.  The absence of $p$-dependence at $U=8$ suggests
that the interaction dependence of the coefficients should be that of Fig.\ 
\ref{fig:koeff55}, independent of the boundary fields $p$.
  
The effect of the boundary potential $p$ on the shift of the positions of
the peaks is much larger than the effect of varying $U$ (compare
Fig.~\ref{fig:mom55p} with Fig.~\ref{fig:mom55}).  Since the fit results
for all three $k$ values agree very well with the Bethe Ansatz conjectures,
the confirmation of Eq.~(\ref{eq:kc}) is even more compelling than it was
for the $U$--dependence.


\section{Summary and Conclusions}
\label{sec:Concl}

We have carried out a detailed comparison between the exact Bethe Ansatz
solution and Density Matrix Renormalization Group calculations for the
one-dimensional Hubbard model with open boundary conditions both with and
without an additional chemical potential at both ends.  A direct comparison
of the ground state energies as well as the density and magnetization at
the ends of the chain has allowed us to estimate the accuracy of the DMRG
on the large system sizes used in this work.

We have then compared the behaviour of the Friedel oscillations in the
local density and local magnetization calculated directly using the DMRG
with Conformal Field Theory predictions for the asymptotic forms for which
the exponents can be calculated using the Bethe Ansatz.  We have performed
this check for two different fillings, $n_e = 0.55$ and $n_e = 0.70$ for
the case without boundary potentials, $p = 0$.  We have obtained results
consistent with the CFT predictions in all cases except those in which it
is clear that the accuracy of the fitting procedure breaks down.  Such a
breakdown occurs when a particular peak in the fourier transform of the
density or magnetization becomes lightly weighted and thus poorly defined.
This occurs principally for the $k_n=(k_{F,\uparrow}+k_{F,\downarrow})$
peak, especially at small $U$ values.  The good agreement between the CFT
forms and BA values of the exponents and the DMRG calculations provides
both a confirmation of the CFT predictions and a way to test the accuracy
of the DMRG and of the effectiveness of fitting procedures for the Friedel
oscillations.

In addition, we have proposed a relation between the $1/L$ corrected mean
values in the density and magnetization and the $1/L$ corrections occuring
in the BAE.  This conjecture is supported by good agreement between mean
values obtained from the fit to the DMRG data and the BAE results.

We have been able to extract for the first time the interaction dependence
of the amplitudes for the Friedel oscillations, a property not possible to
calculate in the framework of the CFT, and have found the correct behaviour
in the $U\to 0$ limit.

Finally, we have turned on boundary chemical potentials at $n_e=0.55$ and
examined the $p$-dependence of the critical exponents, the amplitudes and
our conjectures for the behaviour of the mean density and magnetization.
The different $p$-regimes that we have considered yield qualitatively
distinct Bethe Ansatz solutions that are physically connected with the
formation of different types of bound states at the system boundaries.  In
agreement with BA/CFT predictions, we have found that the critical
exponents are independent of $p$ and that the influence of $p$ on the
amplitudes is very weak.  We have also found that our conjectures for the
$1/L$ corrected mean values of the density, magnetization and the
wave-vector hold in all of the physically different $p$-regimes.

The combination of analytical and numerical methods presented here has
yielded new insights into both.  The success of the numerical techniques
will now allow the examination of more complicated systems that are not
exactly solvable.  Through comparison with the DMRG calculations, we have
also been able to show that more information is contained in the BAE than
is obtained from a direct interpretation via Conformal Field Theory.


\acknowledgments 
We would like to thank J.\ Voit for helpful discussions.
This work has been supported by the Deutsche Forschungsgemeinschaft under
Grant No.\ Fr~737/2--2 (G.B.\ and H.F.) and by the Deutsche
Forschungsgemeinschaft under Grant No.\ Ha~1537/14--1 (B.B.). R.M.N.\ was
supported by the Swiss National Foundation under Grant No.\ 20-46918.96.
\end{multicols}


\widetext 

\begin{multicols}{2}

\end{multicols}
\end{document}